\pdfoutput=1
\documentclass[a4paper,12pt]{article}
\usepackage[utf8]{inputenc}

\usepackage[T1]{fontenc}
\usepackage{mathptmx}

\usepackage{fullpage, graphicx}
\graphicspath{{./Images/}}
\usepackage{siunitx}
\usepackage[backend=biber,sorting=none,style=nature]{biblatex}
\usepackage{url}
\usepackage{float}
\usepackage{multicol}
\usepackage{chemformula}
\usepackage{hyperref}
\usepackage{amssymb}
\usepackage{physics}
\usepackage{mathtools}
\usepackage{indentfirst}
\usepackage{caption}
\usepackage{setspace}
\usepackage[symbol]{footmisc}

\DeclareSIUnit\atm{atm}
\begin{document}
\pagestyle{empty}                       

\begin{center}
        \fontsize{20}{27}\selectfont\bf \uppercase{Nuclear-Elastic Scattering of $\,^9$B\MakeLowercase{e}\,+$^4$H\MakeLowercase{e} using the Proximity Potential}\\         
\end{center}
\begin{center}
        \renewcommand{\thefootnote}{\fnsymbol{footnote}}
        \large Austin A. Morris
        \footnote[2]{Supervised by Alan Shotter.} \\                    

        \vspace{2mm}
        \small\it School of Physics and Astronomy, University of Edinburgh, May 2020                                    
\end{center}
\vspace*{1.5mm}

\begin{spacing}{0.8}
\renewcommand{\baselinestretch}{0.75}
\noindent {\footnotesize\textbf{Abstract}––A nuclear-elastic $^9$Be\,+$^4$He reaction is investigated. A tangential literature overview of the proximity potential is presented, necessitated by an approximate method for calculating $l\text{-values}$. The model is surveyed intensively, from classical conception (liquid drop model) to physical implication (fusion barrier height), leading to a second-part calculation of the $^9$Be\,+$^4$He angular distribution. The results obtained are limited by the exclusion of phase variation, but provide enough accuracy to give order-of-magnitude estimates for nuclei interacting peripherally, in agreement with experiment.}
\end{spacing}




\pagestyle{plain}                               
\setcounter{page}{1}   

\section{Introduction}

This paper is focused upon calculating the angular distribution of scattering Beryllium and Helium nuclei. The $^9$Be\,+$^4$He scattering is assumed to be elastic (no kinetic energy lost) and peripheral (the nuclei barely graze one another). The collision is treated semi-classically with energies ranging from 9.5–20 MeV. For large separation distances, the electrostatic potential repels the positively charged nuclei, whereas for close-range impact, the strong nuclear force tends them toward inelastic processes and fusion. At the surface, a third pseudo-force arises from centrifugal effects generated by angular momentum. The proximity potential is proposed to try to understand the angular distributions, since the reaction is dominated by surface processes.

The angular distributions are thought to suggest that a surface interaction process is dominant, which leads to a specific range of $l\text{-values}$ (angular momenta) for a particular energy.  The angular momentum quantum number $l$ must first be calculated to determine the angular distribution. Suitable values may be calculated by using the proximity potential, which considers two infinite planes of matter interacting between finite separation distance, with corrections for curvature. The model is of mathematical origin, projected through the physical lens of the liquid drop. The first section of this paper may be outlined as such: the classical conception, physical assumptions, and peripheral applications of the proximity potential, comprising a large literature survey.  

The second section of this paper concerns the computational calculation of the angular distribution for the $^9$Be\,+$^4$He interaction using the information about the proximity potential of Section 2. The total differential scattering cross section will be formalized as the sum of the Coulomb and angular momentum distributions, treated semi-classically as to allow interference between probability amplitudes. Then using suitable $l\text{-values}$ approximated by the nuclear proximity potential, the results will be compared with a 1964 paper studying the same reaction. Errors originating from the proximity potential and related assumptions will be discussed, with some commentary on the optical model and overarching limitations.
\\

\section{Proximity Formalism and Application}

Heavy-ion ($A>4$) scattering and fusion processes are of key interest in the current study of nuclear physics \cite{18}. Heavy-ion fusion is particularly important for understanding the creation of neutron-rich super-heavy elements \cite{Dutt}. An accurate method for determining the interaction potential between two nuclei has been a field of strong interest for the past fifty years. The applications of such a potential would be manifold; knowing the nuclear potential value would lend more powerful quantitative predictions in the areas of multi-fragmentation, cluster decay, and particle formation.

Current efforts are being undertaken to better understand complete and incomplete fusion, and super-heavy elements at low-incident energies \cite{Dutt2}. The \textit{proximity potential} is the foundation of many low-incident energy nuclear potential studies \cite{Dutt2},\cite{Shotter}. More than sixteen proximity potential iterations have been made, often from variations of the semi-empirical mass formula parameters used in the calculation of the Myers-Swiatecki mass formula \cite{Dutt2},\cite{Aygun}. The potential is most useful for the calculation of fusion barriers and cross sections, and has been studied for both symmetric and asymmetric nuclei. Designed specifically for ion-ion surface interactions, the value of the potential is further relevant because of the strong surface force influence in stellar fusion events \cite{Shotter}.

Most generally, the proximity potential considers two nuclei interacting within the range of the strong nuclear force \cite{Shotter}. The potential is classically comprised of two components: one based on a geometric factor for the nuclei curvature, and another on the separation factor between two infinite planes of nuclear matter \cite{Shotter},\cite{Ghodsi}. Since the potential is more often used for heavier nuclei, it is unclear whether its application to light nuclei will yield accurate angular momenta estimates.

Before weighing the application to the $^9$Be\,+$^4$He reaction, an extensive survey of the proximity potential will be presented, beginning with the liquid drop model, acquiring shell corrections to satisfy surface energy constraints, and finally formalizing the potential and its different versions.

\subsection{Liquid Drop Model}

Before formalizing the proximity potential for surface interactions, it is worth inspecting the model from which it arises, which will help later to understand the underlying physical assumptions involved, error therein, and areas of possible improvement. 

The \textit{liquid drop model} has long been used for describing the characteristics of nuclei. The basis of the model is to treat the nucleus as a liquid drop composed of nucleons bound by nuclear force. The treatment is quasi-classical, ignoring quantum shell effects, giving a uniform distribution of nucleons in phase space \cite{VM}. The distribution is then inhomogeneous for shell-model corrections, which hold strong dependence on nucleon density at Fermi energy. Along the surface edge, the outer nucleons have surface tension like a liquid drop; the nucleus is spherical in its ground state but may be deformed by added energy \cite{Ludwig}. The charged liquid drop with surface tension is a good model for describing nuclei behaviour, particularly at higher atomic numbers \cite{Swiatecki}.

The relationship of the nucleus to the binding energy in the liquid drop model is given by the semi-empirical mass formula, first described by von Weizsäcker and Bethe \cite{Ludwig}. Problems with the shell-less formulation arise from the nuclear mass dependence on nuclear surface deformations \cite{VM}. It is there that nucleon density falls-off in very short distance of the order $\sim r_0 A^{-\frac{1}{3}}$ fm \cite{VM},\cite{Ludwig}, where $r_0$ is the radius constant and $A$ the atomic mass number. Since the surface and volume tensions are not precisely known, the model application to surface interactions is incomplete without shell corrections.

\subsection{Shell Corrections and Surface Energy}

The nuclear surface may be approximated by a continuous and asymptotic liquid drop with shell corrections \cite{Swiatecki}. The Myers and Swiatecki semi-empirical mass formula (1965) adds a shell modification depending on the strength and position of magic numbers. The obtained mass formula contains four liquid drop parameters of various energies and three shell expressions for bunched nucleons and proton Fermi gases, which vanish as a Gaussian function for large deformations. The Myers-Swiatecki formula is:

\begin{align}
\label{eqn:M}
    M(N,Z,\text{shape}) &= M_{\text{liquid drop}} + M_{\text{shells}}\\
    \nonumber &= M_nN + M_HZ+E_{\text{volume}}+E_{\text{surface}}+E_{\text{Coulomb}}+ \text{corrections}.
\end{align}
\\
The shell correction mass-formula assumes a leptodermous (thin-skinned) nucleus, and is supported by Seyler Blanchard type Thomas-Fermi systems that consider asymmetry and surface diffuseness \cite{Ludwig}. Ludwig \textit{et al}. (1971) updates the formula to thirteen parameters, after accounting for additional surface diffuseness effects.

Treating the liquid drop as a Fermi gas, the surface energy is found to be a function of the nuclear radius, Fermi energy, and surface tension. Using experimental binding energies \cite{77}, the Myers-Swiatecki surface energy coefficient $\gamma\,$ is given by

\begin{align}
\label{eqn:surfacecoeff}
    \gamma = \gamma_{\,0} \left(1 - k_s\,A_s^2 \right)
\end{align}
\\
\noindent for a surface energy constant $\gamma_{\,0} 	\approx 0.9517\,\text{MeV/fm$^2$}$, surface asymmetry constant $k_s \approx 1.7826$, and asymmetry parameter $A_s = \frac{N - Z}{N + Z}$ (accounting for neutron and proton excess of the combined system). Equation (\ref{eqn:surfacecoeff}) has been many times updated for improved calculations of the surface asymmetry and surface energy constants, which arise from the liquid drop model with shell corrections \cite{Dutt}.

\subsection{Surface Potential Formalism}

Blocki \textit{et al}. (1977) first proposed a generalized theorem for relating two curved objects using their inter-atomic potential \cite{18}. Classically formulated, the proximity potential is used to account for heavier atomic nuclei with short-range interactions acting via the strong nuclear force. In addition to the case of two atomic nuclei, the approximation may be also extended to two small cylinders of silicate groups, such as mica \cite{77}.

The shape-dependent potential energy of two curved objects consists of a bulk and surface-layer term, where the decomposition arises when the surface curvature is larger than the surface thickness, conditional upon the system being simply-connected (which a single nucleus is). The approximation may be improved by a curvature correction to the leading area-dependent term in the surface energy, where the necessary conditions are satisfied \cite{77},\cite{Ran}.

For a leptodermous system as previously considered, the total energy of the nucleus may be approximated as the sum of bulk and surface-layer energy terms such that
\begin{equation}
\label{eqn:V-tot}
    V = V_{P} + V_{S},
\end{equation}
\noindent where for low-order curvature corrections (for contorted surfaces) the proximity energy $V_P$ is added to the surface energy $V_S$. For curvatures at points of separation comparable to the diffuseness of the nuclei edges, gap configurations of separation distance $s$ are considered for separated ($s > 0$) and overlapping nuclei ($s < 0$) \cite{79}. 
\\
\begin{figure}[H]
    \hspace*{-2.5mm}
    \centering
    \includegraphics[width=9cm]{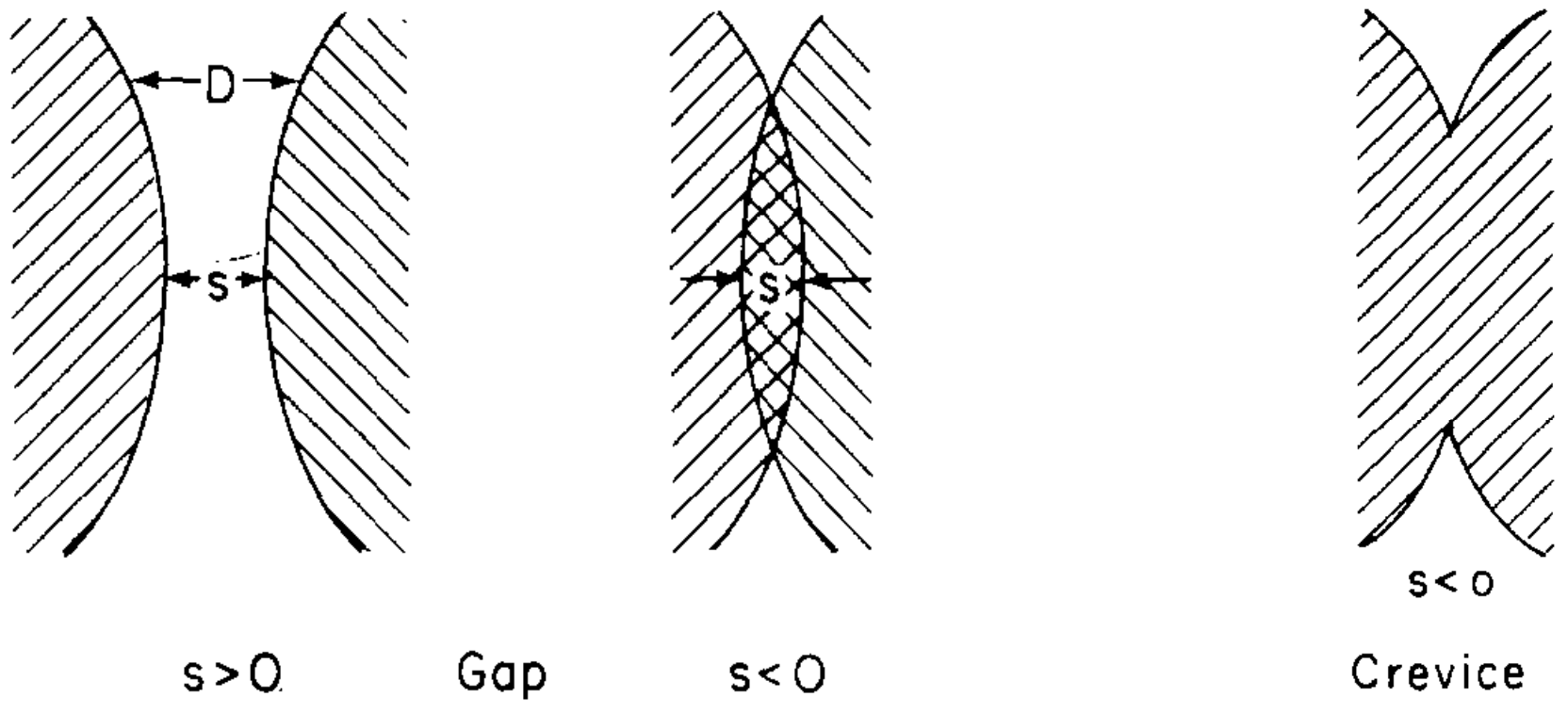}
    \caption{Gaps and crevice diagrams given in reference \cite{79}.}
    \label{fig:Gaps}
\end{figure}

\noindent The proximity energy is equal to the double-integral of the interaction energy of two parallel surfaces of separation $D$, with corrections for gaps and crevices \cite{Ran},\cite{79}:

\begin{align}
\label{eqn:V-P}
    V_P = \iint e(D) \,d\sigma.
\end{align}
\\
\noindent The key trick is the transformation of the two dimensional integral according to gap configuration. Equation (\ref{eqn:V-P}) becomes

\begin{align}
\label{eqn:V-PC}
    V_P = \int_{\text{\textit{s} or 0}} \!\!\!\!\!\!\!e(D) J(D-S)\,dD,
\end{align}
\\
\noindent where the Jacobian of the transformation

\begin{align}
    J = \frac{d\,(\text{area in x-y plane})}{dD} = \frac{d\sigma}{dD}
\end{align}
\\
\noindent is the gap width distribution \cite{79}. The surface-layer term in (\ref{eqn:V-tot}) becomes

\begin{align}
\label{eqn:V-sl}
    V_{S} = \iiint{\Theta_S} = \gamma \iint_\Sigma \,d\sigma
\end{align}
\\
\noindent for a surface energy function defined as
\begin{align*}
    \Theta_S \coloneqq \eta - a_B\rho.
\end{align*}
Equation (\ref{eqn:V-sl}) gives an integration over the surface energy, where the integral over $\eta$ is the actual energy and the integral over $a_B\rho$ is the energy that the same amount of matter would require in the bulk term. This is equal to the double-integral on surface $\Sigma$ over the infinitesimal area element $d\sigma$ of the surface energy coefficient $\gamma$.

For a strictly leptodermous system in which the electrostatic interaction is considered, the surface energy fluctuates due to non-constant equilibrium bulk densities according to the liquid drop model \cite{77}. Indeed, equation (\ref{eqn:V-tot}) can be imagined in the context of the liquid drop, whereby strongly-interacting particles with short and long mean-free paths may be treated by classical mechanics and quantum mechanics, respectively.

\subsection{Nuclear Proximity Potential}

Using the mathematical formalism of Blocki \textit{et al}. (1977) and physical assumptions of the liquid drop model, the proximity potential may be further developed for heavy-ion interactions \cite{Ghodsi}. The interface between macro-surfaces may be transformed to nuclear surfaces, hence the nuclear potential may be defined as the proximity potential previously formalized: $V_N \equiv V_P$, where for a paraboloidal approximation \cite{Ran}, equation (\ref{eqn:V-PC}) becomes

\begin{align}
\label{eqn:V_N}
    V_{N}(r) = 4\pi \gamma \, b\bar{R}\, \Phi (\xi) \quad \text{[MeV]}.
\end{align}
\vspace{1mm}
\\
\noindent The $\gamma\, b\bar{R}$ term describes the shape and curvature of the nuclei and the universal function $\Phi (\xi)$ describes the minimum separation distance $s$ of two parallel nuclear planes \cite{Shotter},\cite{Ghodsi}. The surface energy $\gamma\,$ is based on the Myers-Swiatecki formula for the shell-model liquid drop \cite{Swiatecki},\cite{77}. The reduced radius $\bar{R}$ is given by the matter central radius $C_i$ \cite{Ghodsi}:

\begin{align}
\label{eqn:reduced_radius}
    \bar{R} = \frac{C_1\,C_2}{C_1+C_2},
\end{align}
\noindent where
\begin{align}
\label{eqn:central_radius}
    C_i = R_i\left[1-\left(\frac{b}{R_i}\right)^2 + \dots\right] \quad \text{for } i=1,2.
\end{align}
\noindent The effective sharp radius is
\begin{align}
\label{eqn:sharp_radius}
    R_i = 1.28A_i^{1/3} - 0.76 + 0.8A_i^{-1/3} \; \text{[fm]} \quad \text{for } i=1,2. 
\end{align}
The dimensionless universal function $\Phi (\xi=s/b)$ depends only on separation distance $s = r -C_1 - C_2$ fm (Figure \ref{fig:Gaps}) and is derived using Seyer-Blanchard type Thomas-Fermi nucleon-nucleon interactions for a leptodermous system \cite{Ghodsi},\cite{Ludwig},\cite{77}.

\subsection{Models and Limitations}

Over the last fifty years there have been more than sixteen proximity potentials proposed \cite{Aygun}. Most models are generally composed of similar surface energies and universal functions, but with refined values, such as Proximity 1988, which varies only in the surface energy and asymmetry constants of equation (\ref{eqn:surfacecoeff}). 

Bass 1980 calculates a nuclear proximity potential from experimental fusion cross sections \cite{Dutt2}. The potential is formulated as:

\begin{align}
\label{eqn:bass}
    V_{N}(r) = -\frac{R_1\,R_2}{R_1+R_2} \Phi (\xi)=(r-R_1-R_2) \quad \text{[MeV]},
\end{align}
\noindent where 
\begin{align}
\label{eqn:bass_radius}
    R_i = R_{0i}\left(1-\frac{b^2}{R_{0i}^2} \right) \quad \text{for } i=1,2.
\end{align}
The surface diffuseness $b^2 = 0.98$ fm$^2$ is again taken to be close to unity. Similar to the effective sharp radius of Proximity 1977 (\ref{eqn:sharp_radius}),
\begin{align}
\label{eqn:bass_radius2}
    R_{0i} = 1.28A_i^{1/3} - 0.76 + 0.8A_i^{-1/3} \; \text{[fm]} \quad \text{for } i=1,2. 
\end{align}
\noindent The main difference between the two potentials arises from the calculation of the universal function, which considers slightly different gap configurations \cite{Aygun}.

One of the more accurate approximations, Winther 1995 uses a Woods-Saxon folding-potential calculated from the densities of two nuclei and their effective two-body force, with parameters adjusted to fit elastic scattering data \cite{Dutt2},\cite{AW}:

\begin{align}
\label{eqn:AW}
    V_{N}(r) = -\frac{V_0}{1+\exp\left(\frac{r-R_1-R_2}{a}\right)} \quad \text{[MeV]},
\end{align}
\noindent where
\begin{align}
\label{eqn:AW_V_0}
    V_0 = 16\pi \frac{R_1\,R_2}{R_1+R_2} \gamma \,a
\end{align}
\noindent and
\begin{align}
\label{eqn:AW_radius}
    R_{i} = 1.2A_i^{1/3} - 0.09 \;\; \text{[fm]} \quad \text{for } i=1,2. 
\end{align}
\\
The surface energy coefficient is reformulated for projectile-target nuclei as 

\begin{align}
\label{eqn:AW_surface}
    \gamma = \gamma_{\,0} \left[1-k_s\left(\frac{N_p - Z_p}{A_p}\right)\left(\frac{N_t - Z_t}{A_t}\right)\right],
\end{align}
\\
\noindent where $\gamma_{\,0} \approx 0.95$ MeV/fm$^2$, $k_s \approx 1.8$, and surface diffuseness parameter
\begin{align}
\label{eqn:AW_diffuseness}
    a = \left\{\frac{1}{1.17\left[1+0.53\left(A_1^{-1/3}+A_2^{-1/3}\right)\right]}\right\} \quad \text{[fm]}.
\end{align}
\\
\noindent The AW95 proximity potential is designed for elastic surface reactions, but also gives good results for Coulomb excitation (inelastic scattering) \cite{AW}.
\\

The \textit{optical model} should be mentioned, though it is not based on any of the proximity potentials, nor is it catered to surface interactions. The model is primarily used for the calculation of the angular distribution in the Taylor (1965) scattering data, which will largely consume the second section of this paper. The optical model potential is

\begin{align}
\label{eqn:OM}
    V(r) = V_C(r)-\left[U(r)+iW(r)+iW'(r)-V_{so}(r)\,\pmb{I \!\cdot \!l}\right],
\end{align}
\\
where $U(r)$ is the depth of the optical well, $W(r)$ and $W'(r)$ are imaginary volume and surface variables, $V_{so}(r)$ is the spin-orbit part of the potential, and the $\pmb{I \!\cdot \!l}$ term is a spin-coupling correction for determining $l\text{-values}$ at the potential barrier \cite{Taylor}. The potential terms are scaled by diffuseness parameters, and the model falls off as separation $r$ approaches infinity, mirroring the density distribution. The model solves the nucleus-nucleus Hamiltonian by distorting incident and outgoing waves to find the transition probability (distorted-wave Born Approximation). Limitations to this approach arise when angular momenta processes dominate surface collisions, such as for back-angle scattering.

The proximity potential may similarly be used to calculate $l\text{-values}$ but with a considerable amount more simplicity. Limitations in the different proximity potentials arise from the (i) nuclei radius, (ii) universal function, and (iii) surface energy coefficient \cite{Ghodsi}. (i) varies outward with isospin and mass, and (iii) with surface diffuseness \cite{Dutt}. The estimation of (iii) may be further improved by more accurate cross section data. The original proximity formalism (\ref{eqn:V_N}) overestimated fusion-barrier heights by $\sim 4\%$ and other experimental data by $\sim 7\%$ for symmetric nuclei based on the Myers-Swiatecki surface energy coefficient $\gamma$. The next section will explicitly use Proximity 1977 to calculate suitable $l\text{-values}$ in an attempt to better understand the $^9$Be\,+$^4$He interaction.

\newpage

\section{Elastic Scattering of $^9$Be\,+$^4$He Nuclei}

To investigate the effectiveness of the proximity potential for light-nuclei, an elastic scattering event for Beryllium and an incident alpha particle is considered. As will be seen, the angular distribution for such an event is dominated by angular momenta, evidenced by a rise in the backward direction that favors orbital phenomena at the surface \cite{Shotter}. For this reason, it has been suggested that the proximity potential be used to understand the angular distributions, since the potential is designed specifically for surface interactions. 

In order to determine suitable $l\text{-values}$, the Coulomb, orbital, and nuclear potentials will be formalized and summed. Upon determining the angular relationship for intermediate energies 9.5 and 20 MeV, the respective angular distributions will be considered. Using data from Taylor \textit{et al.} (1965) \cite{Taylor}, back-angle scattering will be analyzed for higher scattering angles, with specific relevance to elastic grazing. Commentary on the optical model will necessarily follow the comparisons made with the Taylor distributions, and a discussion on the performance of the proximity potential will conclude this section.

\subsection{Direct Reactions in the Entrance Channel}

\begin{figure}[H]
    \hspace*{-7mm}
    \centering
    \includegraphics[width=9.5cm]{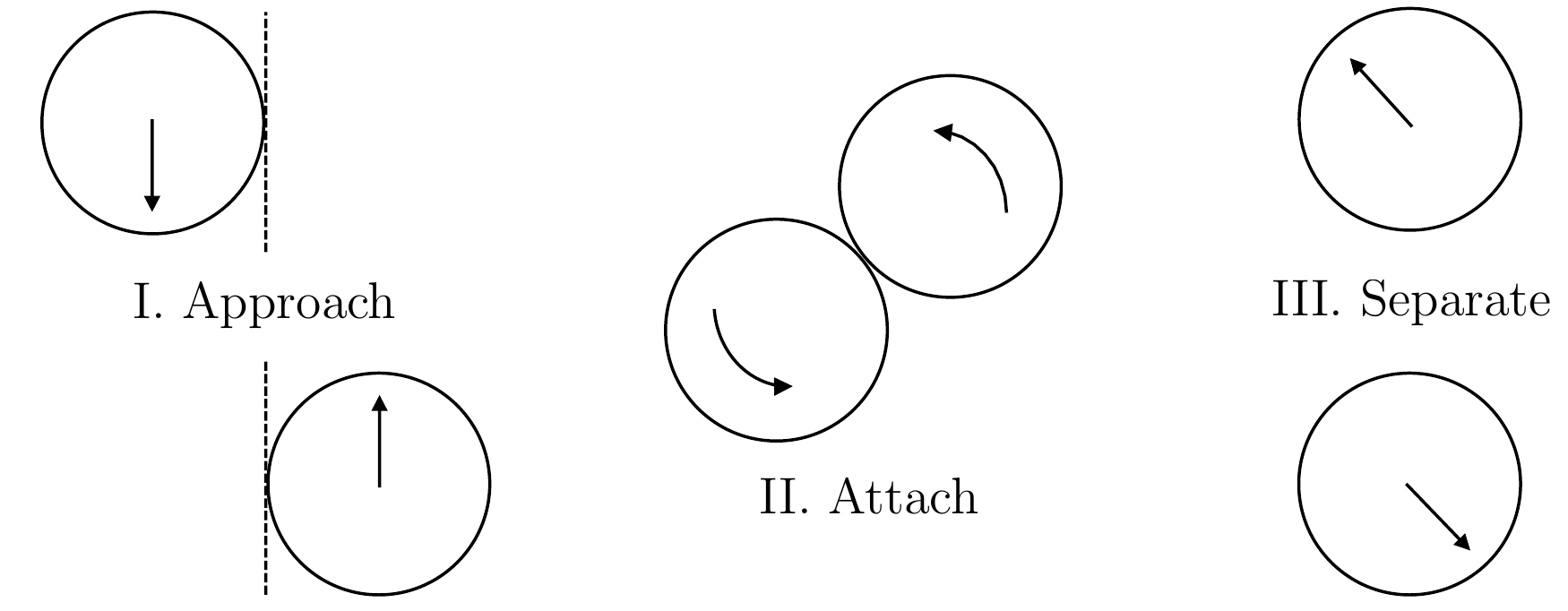}
    \caption{Entrance channel process for two grazing nuclei depicted in reference \cite{Shotter}.}
    \label{fig:ECM}
\end{figure}

The $^9$Be\,+$^4$He surface scattering event involves two charged, light nuclei. The peripheral interaction (direct reaction) will assume negligible inelastic contributions \cite{Krane}. The proximity potential is of related interest, since angular momenta in surface reactions are generally larger than in central reactions \cite{Shotter}. Large angular momentum $l\text{-values}$ lower the probability of two nuclei fusing, but increase the likelihood of strong rotational effects. In the entrance channel model, the direct reaction is well described by a dynamic (I) approach, (II) attachment, and (III) separation process (Figure \ref{fig:ECM}). It is at the contact point in (II) that a pseudo-force provides centrifugal influences strong enough to generate orbital motion. For elastic processes the nuclei break apart (III), showing a steep rise in cross section at backward scattering angles. To understand this phenomena, the three acting potentials, Coulomb, orbital, and nuclear (proximity), must be further generalized.
\\

\subsection{Surface Potentials}

\begin{figure}[H]
    \centering
    \includegraphics[width=9cm]{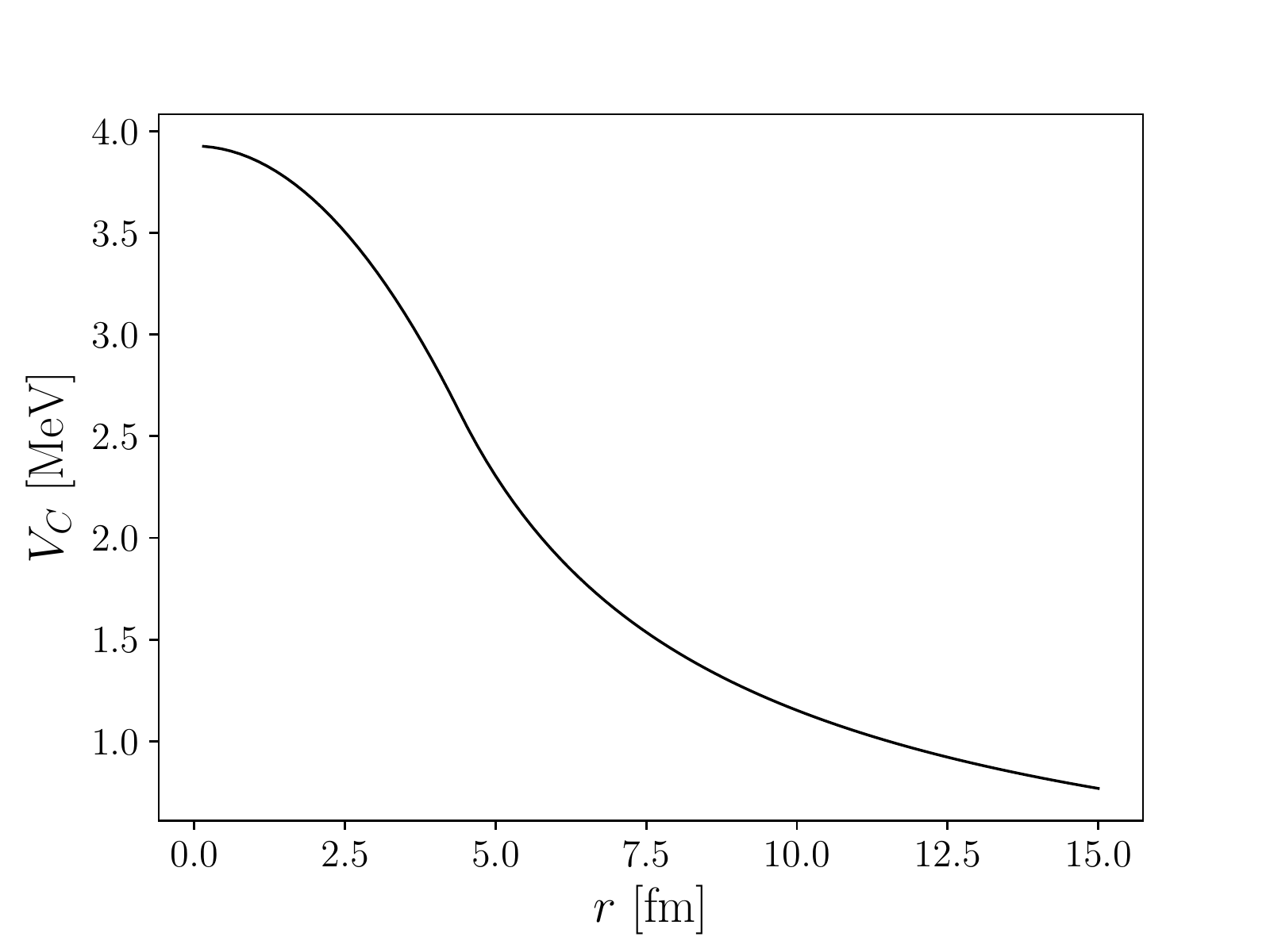}
    \caption{Isolated Coulomb potential for $^9$Be\,+$^4$He elastic scattering.}
    \label{fig:Coulomb_Potential}
\end{figure}

For spherically symmetric charge distributions, the Coulomb potential for projectile-target nuclei is
\begin{align}
\label{eqn:V_C}
    V_{C}(r) = Z_p\,Z_t\,e^2 \begin{cases}
    \frac{1}{r} & \text{for $r \geq R_c$} \\ 
    \frac{1}{2 R_c} \left[3 - \left(\frac{r}{R_c}\right)^2 \right] & \text{for $r < R_c$} \\ 
    \end{cases},
\end{align}
\\
\noindent where $r$ is the distance between the centers and $R_c$ the radial separation (Coulomb radius) between them \cite{Aygun}. The impact parameter $b$ is roughly the sum of the projectile and target radii:
\begin{equation}
\label{eqn:impactl}
    b \approx R_c = r_p + r_t = r_0\sum_{i=1}^2 A_i ^{\sfrac{1}{3}}
\end{equation}
\\
\noindent for nucleon radius $r_0 \sim 1.25$ fm and nucleon number $A$. Then it follows that for $^9$Be\,+$^4$He scattering $b \approx 4.58$ fm. Figure \ref{fig:Coulomb_Potential} shows the singular Coulomb contribution for this parameter. 

 Orbital angular momentum is conserved in the elastic scattering process: $\,\Vec{p_\theta} \times \Vec{r} = l \hslash$. Solving directly the system Hamiltonian, the rotational potential is

\begin{align}
\label{eqn:V_l}
    V_{l}(r) = \frac{l(l + 1)}{2 \mu r^2} \hslash^2
\end{align}

\noindent for orbital angular momentum quantum number $l$ and reduced mass $\mu = \frac{M_p \, M_t}{M_p + M_t}$ \cite{Shotter}.

\begin{figure}[H]
    \centering
    \includegraphics[width=9cm]{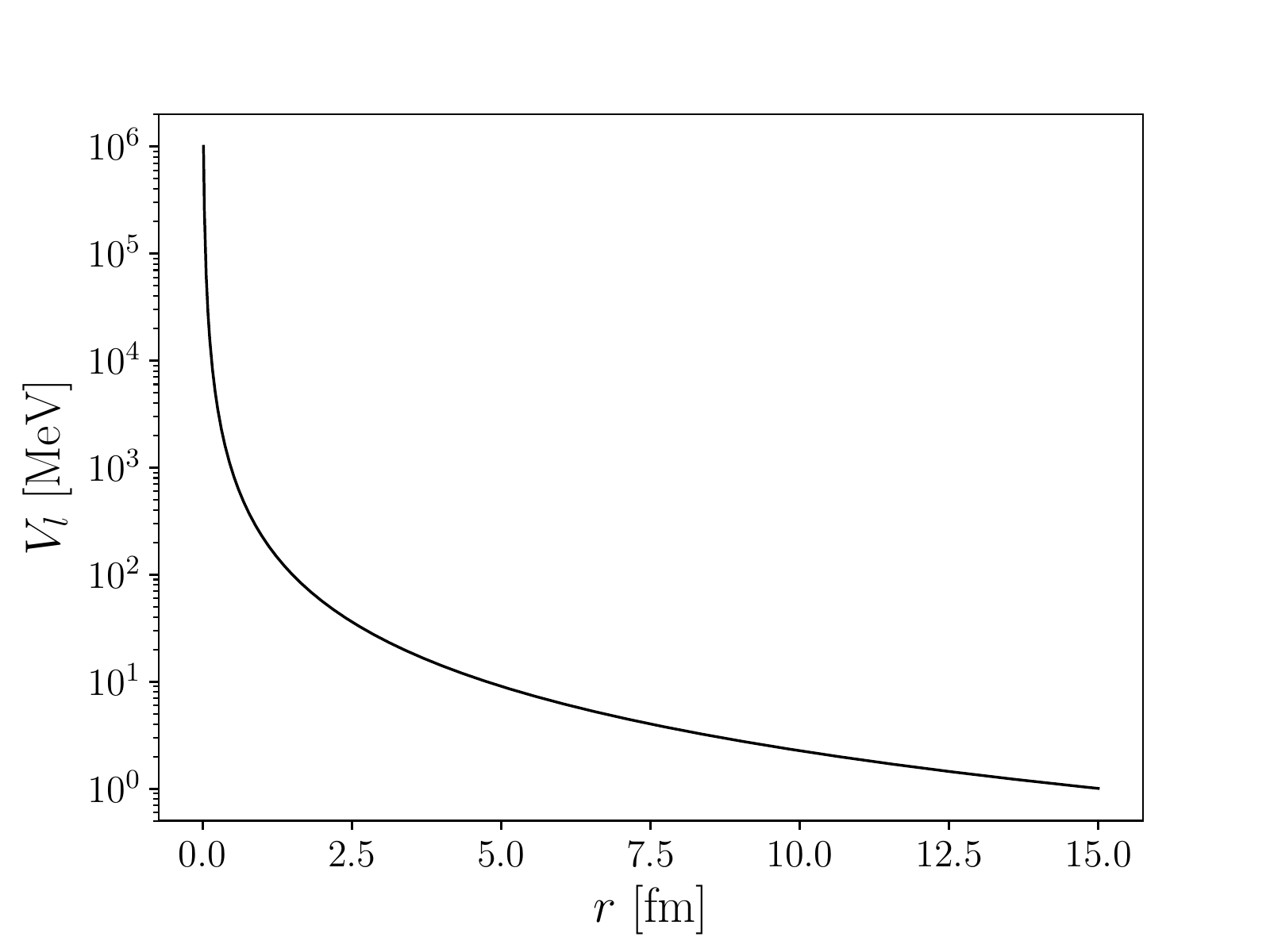}
    \caption{Isolated orbital angular momentum potential for $l=5$ $^9$Be\,+$^4$He elastic scattering.}
    \label{fig:Orbital_Potential}
\end{figure}

 Semi-classically, the impact parameter $b$ is related to $l$ by $p\:\!b = l\hslash$ for translational momentum $p$ and quantized angular momentum $l\hslash$ \cite{Krane}. The discrete value of $l$ may be approximated by
\begin{align}
\label{eqn:lapprox}
    l = {\sqrt{2\mu E}}\, \frac{b}{\hslash},
\end{align}
\\
which gives $l \sim 5$ for 9.5 MeV and $l \sim 7$ for 20 MeV. Figure \ref{fig:Orbital_Potential} shows the orbital potential contribution of an $l=5$ system.
\vspace{-4mm}

\begin{figure}[H]
    \centering
    \includegraphics[width=9cm]{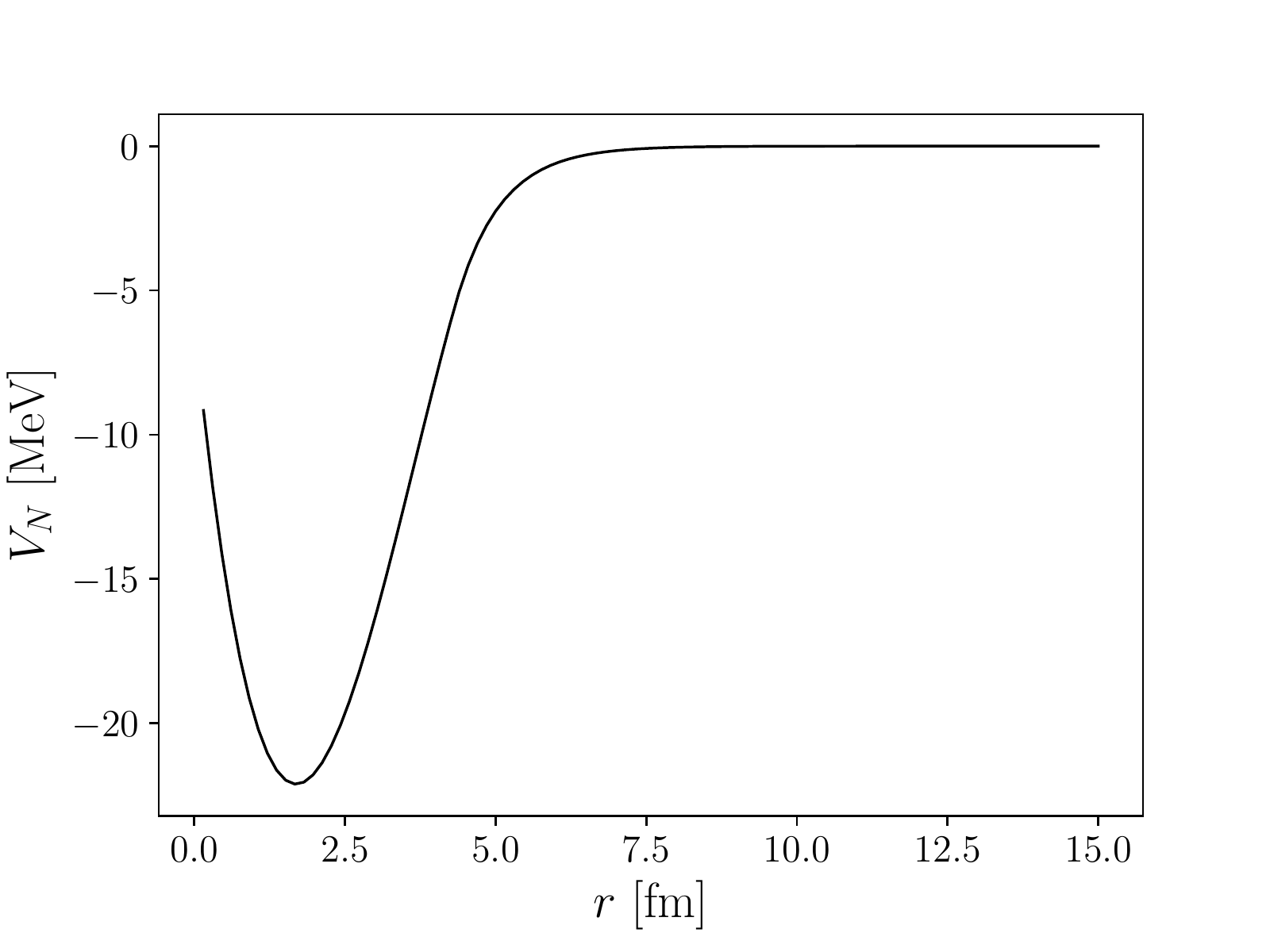}
    \caption{Isolated nuclear proximity potential for $^9$Be\,+$^4$He elastic scattering.}
    \label{fig:Nuclear_Potential}
\end{figure}

Proximity 1977 (\ref{eqn:V_N}) is now used to calculate the strong nuclear surface potential for the $^9$Be\,+$^4$He interaction. The universal function is parameterized by the piecewise function \vspace{2mm}
\begin{align}
\label{eqn:universal}
    \Phi (\xi) = \begin{cases}
    -\frac{1}{2}(\xi-2.54)^2-0.0852(\xi-2.54)^3 & \text{if $\xi \leq 1.2511$} \\ 
    -3.437 \exp \left(-\frac{\xi}{0.75} \right) & \text{if $\xi > 1.2511$} \\ 
    \end{cases}
\end{align}

\noindent for separation distance $s=r-C_1-C_2$ fm and minimum (dimensionless) separation distance $\xi = \frac{s}{b}$ \cite{77}. The width of nuclear surface $b \approx 1$ and the $^9$Be\,+$^4$He ($N=7$, $Z=6$ combined system) asymmetry parameter $A_s \approx 0.0769$. The surface energy coefficient $\gamma$ and mean curvature radius $\Bar{R}$ are calculated using the Myers-Swiatecki formula (\ref{eqn:surfacecoeff}) and reduced radius (\ref{eqn:reduced_radius}). It follows that the shape component is constant and the total potential varies only as a function of $r$, independent of mass. The isolated nuclear proximity potential is plotted in Figure \ref{fig:Nuclear_Potential}.
\vspace{-3mm}

\subsection{Effective Potential for Suitable $l\text{-values}$}

The total interaction potential is the sum of the three acting force components \cite{Ghodsi}. Equation (\ref{eqn:V-tot}) becomes
\begin{align}
\label{eqn:V}
    V(r) = V_C + V_l + V_N.
\end{align}

\begin{figure}[H]
    \centering
    \includegraphics[width=9cm]{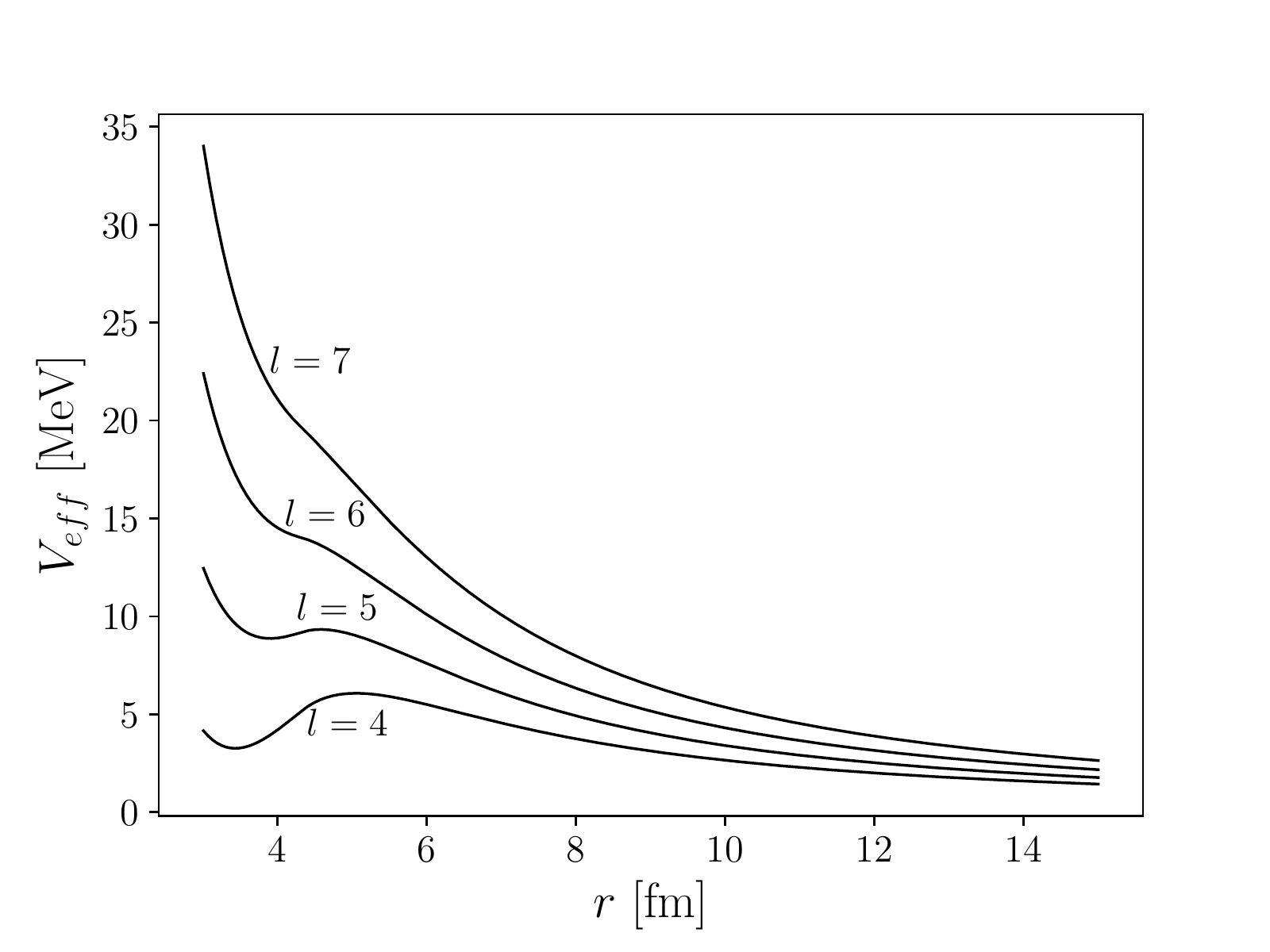}
    \caption{Effective potential $l\text{-values}$ for $^9$Be\,+$^4$He elastic scattering.}
    \label{fig:Effective_Potential}
\end{figure}

Combining the results of Figures \ref{fig:Coulomb_Potential}, \ref{fig:Orbital_Potential}, and \ref{fig:Nuclear_Potential} the effective surface  potential is calculated for a range of $l\text{-values}$ (Figure \ref{fig:Effective_Potential}).

\noindent It is observed that close to the nucleus, the alpha particle goes up the potential barrier as the radial velocity approaches zero, hence there is some rotation around the $^9$Be nucleus as linear kinetic energy is converted into rotational energy. It is assumed that for strong angular momenta effects, the $l\text{-value}$ at the top of the barrier is dominant. It can then be seen that along the Coulomb radius $l \sim 5$ for 9.5 MeV and $l \sim 7$ for 20 MeV, in agreement with equation (\ref{eqn:lapprox}).

\subsection{Angular Distribution Formalism}

Having now calculated suitable $l\text{-values}$ for the mid-size energies of the $^9$Be\,+$^4$He reaction, the angular distributions may be found using the Coulomb and nuclear scattering cross sections.

\begin{figure}[H]
    \centering
    \includegraphics[width=9cm]{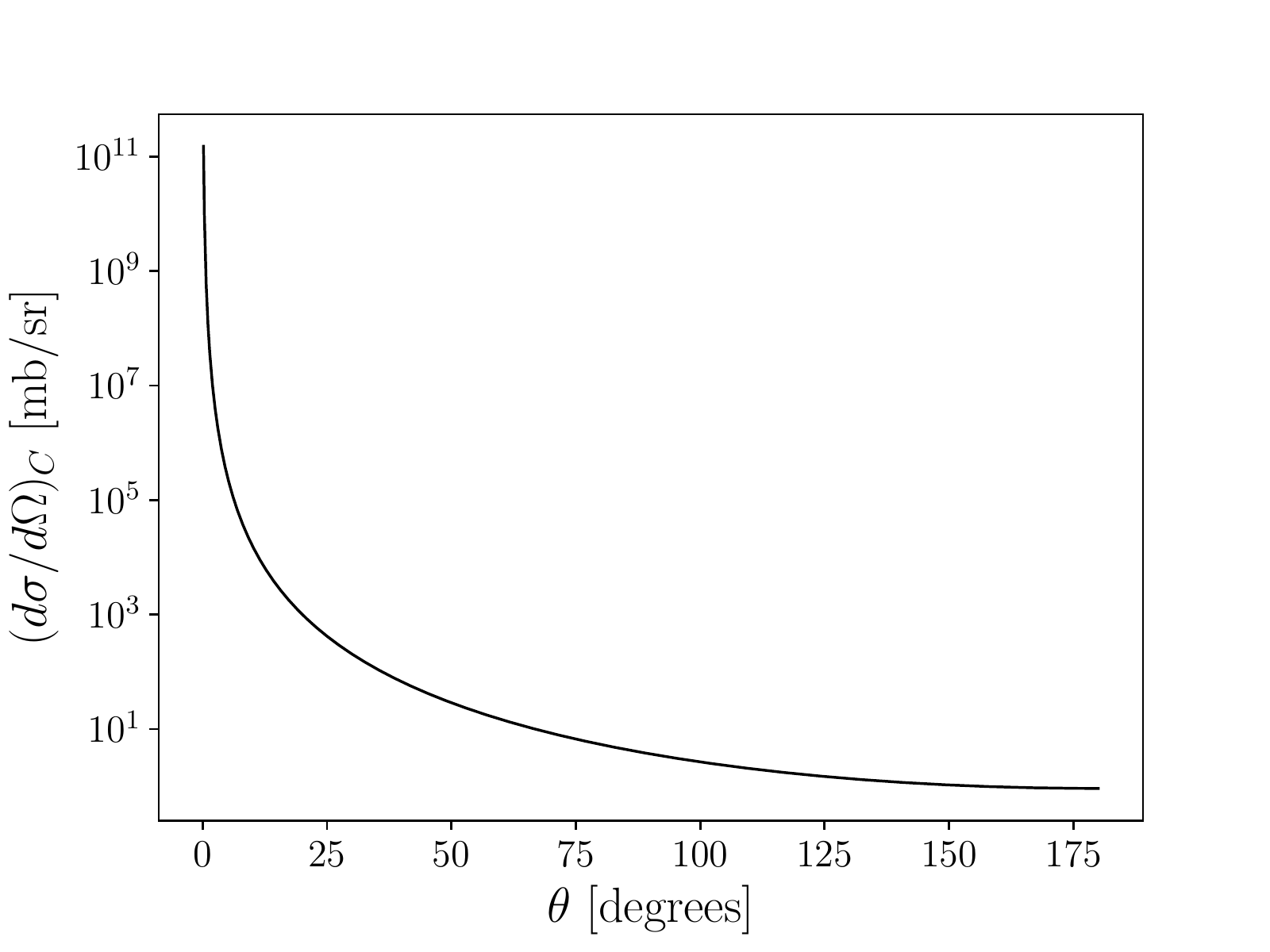}
    \caption{Isolated Coulomb angular distribution for $E=9.5$ MeV $^9$Be\,+$^4$He elastic scattering.}
    \label{fig:Coulomb_Angular_Distribution}
\end{figure}

Considering first the Coulomb interaction, the Rutherford differential cross section is
\begin{align}
\label{eqn:rutherford}
    \left(\frac{d\sigma}{d\Omega}\right)_{\hspace{-1mm}C}=\left(\frac{Z_{\text{Be}}\,Z_{\text{He}}\,e^2}{4\pi\epsilon_0}\right)^2 \left(\frac{1}{4E}\right)^2 \sin^{-4}{\left(\frac{\theta}{2}\right)}
\end{align}
\\
\noindent where $Z\,e^2$ is the nuclear charge, $E$ is the projectile energy, and $\theta$ is the scattering angle \cite{Krane}. The consequent formula is simplified using the dimensionless fine structure constant
\begin{align*}
    \alpha = \frac{1}{4\pi\epsilon_0}\frac{e^2}{\hslash c} \simeq \frac{1}{137.036}
\end{align*}

\noindent so that (\ref{eqn:rutherford}) becomes
\begin{align}
\label{eqn:rutherford2}
    \left(\frac{d\sigma}{d\Omega}\right)_{\hspace{-1mm}C}=\left(\frac{Z_{\text{Be}}\,Z_{\text{He}}\, \alpha \, \hslash c}{4E \sin^2{\left(\sfrac{\theta}{2}\right)}}\right)^2,
\end{align}
\\
where $\hslash c \sim 197.33$ MeV$\cdot$fm, $Z_{\text{Be}}=+4$ and $Z_{\text{He}}=+2$, and $0 < \theta < \pi$. Considering non-relativistic energies between 9.5 and 20 MeV, the angular distribution for the Coulomb effect alone is calculated and plotted (Figure \ref{fig:Coulomb_Angular_Distribution}). It is observed that the Rutherford cross section decreases for higher energies and larger angles.

 The more difficult component of the angular distribution arises from the nuclear effects generated by orbital angular momentum. Considering incident (spinless) nuclei, the $l$ and magnetic quantum number $m$ values may only take discrete values: $l = 0,\: 1,\: 2,\: \dots$ and $m = 0,\: \pm{1},\: \dots,\: \pm{l}$. The wave function $\psi$ of the system is the product of a radial function and angular component given by the spherical harmonics $Y_{lm}(\theta,\phi)$, where
 \begin{align*}
        Y_{lm}(\theta,\phi) = (-1)^m \sqrt{\frac{2l+1}{4\pi}\frac{(l-m)!}{(l+m)!}} \, P_{l}^m (\cos{\theta}) \, e^{im\phi}{,}
\end{align*}
\vspace{-3mm}
\begin{align*}
        \text{and } \;P_l (\cos{\theta}) = \frac{1}{2^l l!} \, \left(\frac{d}{d(\cos{\theta})}\right)^l \left(\cos^2\theta - 1\right)^l \text{ is the Legendre polynomial.}
\end{align*}

\noindent For the $^9$Be\,+$^4$He scattering process, normalization is chosen such that $m = 0$, since integrating over the azimuthal angle gives vanishing contribution:
\begin{align*}
    \int_0^{2\pi}\!\! e^{im\phi}\,d\phi = 0 \quad \text{for all } m = \{n \in \mathbb{Z^+}\}.
\end{align*}
The nuclear differential cross section is proportional to the modulus of the scattering amplitude squared. Since at the top of the potential barrier, the radial energy approaches zero, it can be inferred without explicit calculation of the transition matrix that $\left(\frac{d\sigma}{d\Omega}\right)_{\:\!\!\!N}\propto \left|Y_{l}^0(\theta)\right|^2$. For an elastic scattering process with no phase change, this may be approximated by using a normalization constant:
\begin{align}
\label{eqn:difffl}
    \left(\frac{d\sigma}{d\Omega}\right)_{\hspace{-1mm}N} =  \left|N_{r} \, Y_{l}^0(\theta)\right|^2 \quad \text{for } N_r = \{\lambda + i\omega \in \mathbb{C}\}.
\end{align}

\begin{figure}[H]
    \vspace{-0.1cm}
    \centering
    \includegraphics[width=5.5cm]{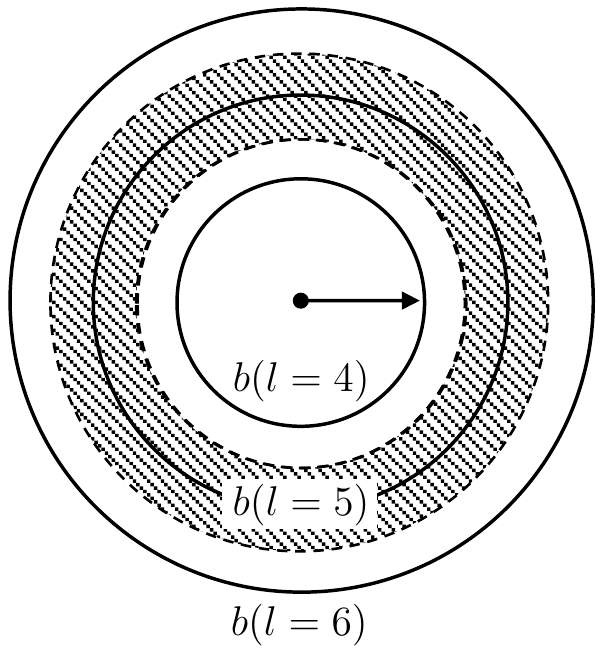}
    \caption{Approximation of the total cross section using $l\text{-values}$ in-between quanta.}
    \label{fig:cross}
\end{figure}

For a given $l\text{-value}$, the scattering area may be divided into fixed radii of varying impact parameters (smooth values). The total cross section may be estimated to be within the range of annuli given between parameters (Figure \ref{fig:cross}), such that 

\begin{align}
\label{eqn:totalcross}
    \pi \left(b_{l+\sfrac{1}{2}}^2 + b_{l-\sfrac{1}{2}}^2 \right) \sim \sigma_T \leq \pi \left(b_{l+1}^2 - b_l^2 \right).
\end{align}
\noindent A relationship between the total and differential cross sections is obtained by integrating over the solid angle:

\begin{align*}
    \sigma_T = \oint_{4\pi} \frac{d\sigma}{d\Omega} \, d\Omega = \int_0^{2\pi}\int_0^\pi \frac{d\sigma}{d\Omega} \sin{\theta} \, d\theta \, d\phi &= 2\pi \int_0^\pi \left|N_r Y_l^0(\theta) \right|^2 \sin{\theta}\, d{\theta}\\
    \\
    &= -\frac{2l+1}{2} |N_r|^2 \int_{1}^{-1} \left|P_l(z) \right|^2 \, dz \\
    &= \sqrt{\frac{2l+1}{2}} |N_r|^2.
\end{align*}
\noindent It follows that the normalization constant squared is the normalized Legendre polynomial times the total cross section:
\begin{align}
\label{eqn:norm}
    |N_r|^2 = \sqrt{\frac{2}{2l+1}} \sigma_T = \|P_l\| \sigma_T \quad \text{and} \quad |N_r| = \sqrt{\|P_l\| \sigma_T} = |\lambda + i\omega|.
\end{align}

\begin{figure}[H]
    \centering
    \includegraphics[width=9cm]{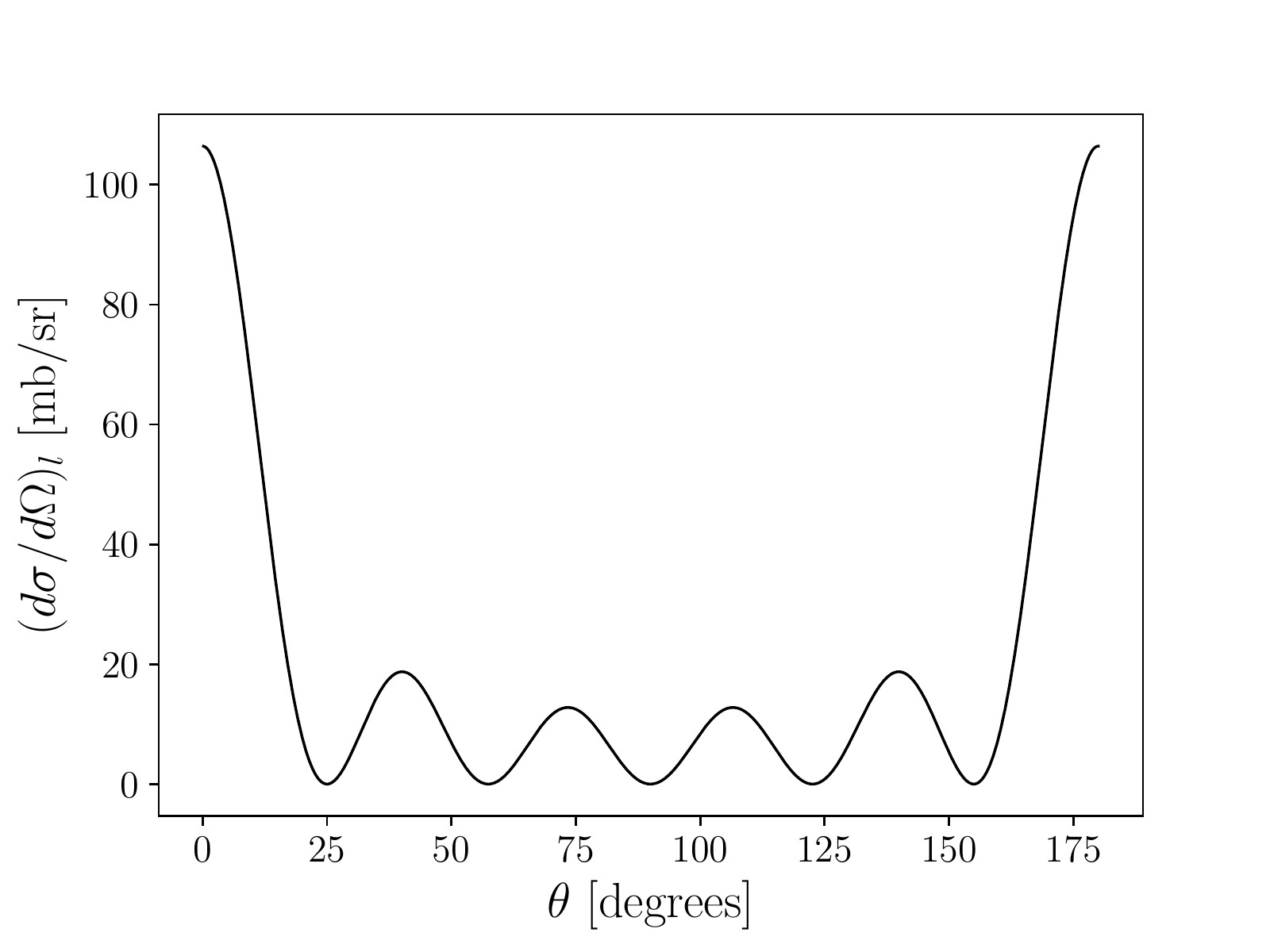}
    \caption{Isolated nuclear angular distribution for $l=5$ elastic scattering.}
    \label{fig:Nuclear_Angular_Distribution}
\end{figure}

From equation (\ref{eqn:difffl}) the angular distribution for the nuclear component is calculated and plotted for the $l=5$ case (Figure \ref{fig:Nuclear_Angular_Distribution}).
\vspace{-3mm}

Combining (\ref{eqn:rutherford2}) and (\ref{eqn:difffl}), the angular distributions can be formalized by scattering amplitudes: 
\begin{align*}
    \left(\frac{d\sigma}{d\Omega}\right)_{\hspace{-1mm}C}=|f_C|^2 \quad \text{and} \quad \left(\frac{d\sigma}{d\Omega}\right)_{\hspace{-1mm}N}=|f_N|^2.
\end{align*}

\noindent The total differential cross section (with interference) is then given by

\begin{align}
\label{eqn:angular_total1}
    \frac{d\sigma}{d\Omega}=|f_C+f_N|^2
    =\left|\frac{Z_{\text{Be}}\,Z_{\text{He}}\, \alpha \, \hslash c}{4E \sin^2{\left(\sfrac{\theta}{2}\right)}} + N_{r} \, Y_{l}^0(\theta) \right|^2.
\end{align}
\\

\subsection{Comparison with Taylor Data}

Taylor \textit{et al.} (1965) provides an extensive investigation of $^9$Be\,+$^4$He elastic scattering data for energies between 4 and 20 MeV \cite{Shotter}. Using real well depths, eight excitation functions and sixteen angular distributions are found \cite{Taylor}, where the optical model potential (\ref{eqn:OM}) is used to calculate the latter. Without the \pmb{$I \!\cdot \!l$} spin-coupling term, imprecise back-angle predictions are made. Bumps in the excitation spectra give strong evidence for rotational processes at the surface \cite{Shotter}. 

Unlike the optical model used by Taylor, the proximity potential is specifically designed for surface collisions between two nuclei. Using the closest $l\text{-values}$ (Figure \ref{fig:Effective_Potential}), the angular distribution for the $E=9.5$ MeV and $E=20$ MeV cases is calculated using equation (\ref{eqn:angular_total1}). The comparative results are produced below:
\\
\begin{figure}[h!]
\hspace{-1.1cm}
    \centering
    \begin{minipage}{0.8\textwidth}
        \centering
        \includegraphics[width=0.9\textwidth]{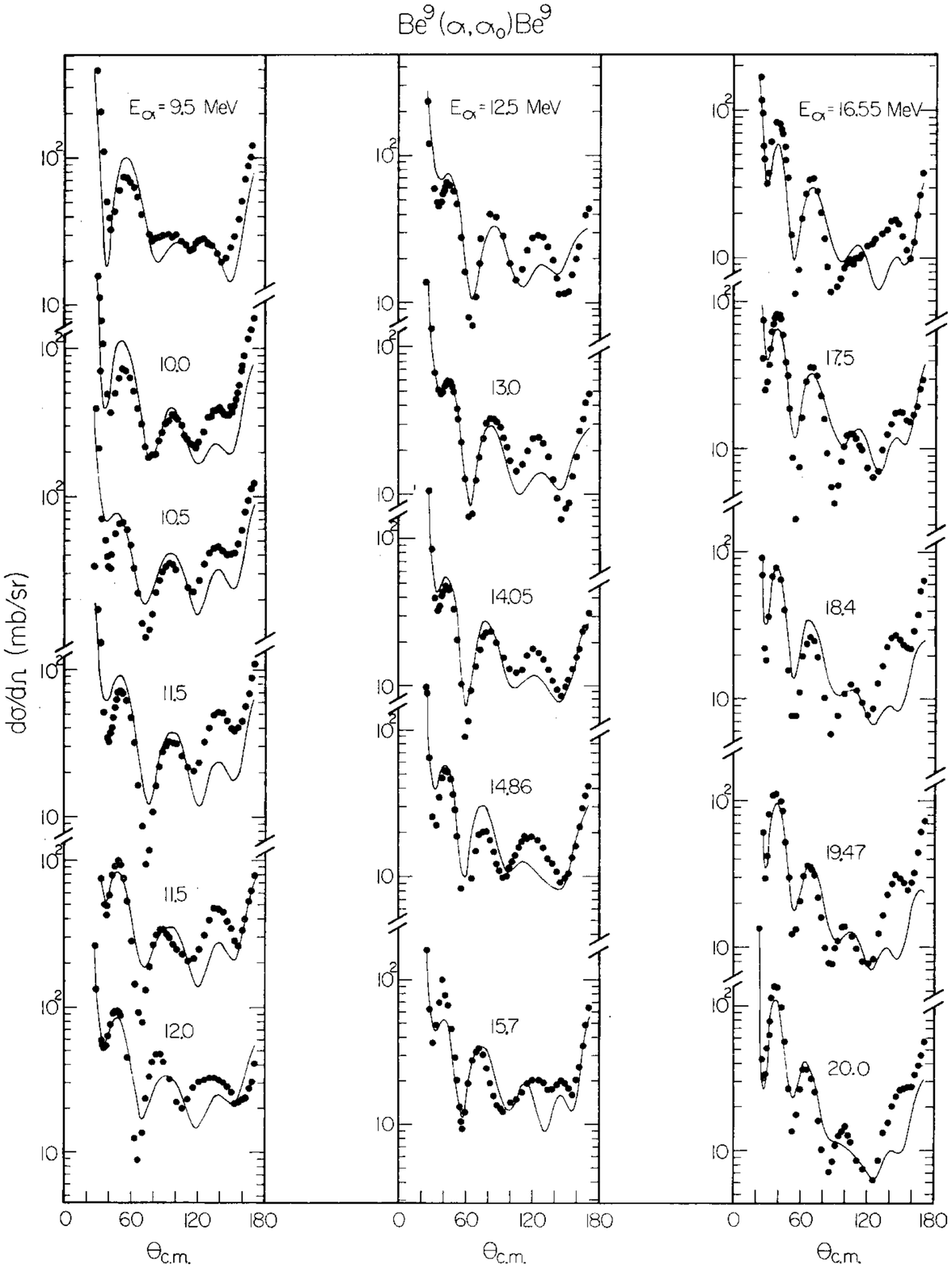}
        \caption{Taylor (1965) elastic scattering data from reference \cite{Taylor}.}
        \label{fig:Taylor_Data}
    \end{minipage}%
    \begin{minipage}{0.25\textwidth}
        \begin{figure}[H]
        \vspace{-0.9cm}
        \hspace{-0.35cm}
            \hspace{-1cm}
            \centering  \includegraphics[width=4cm]{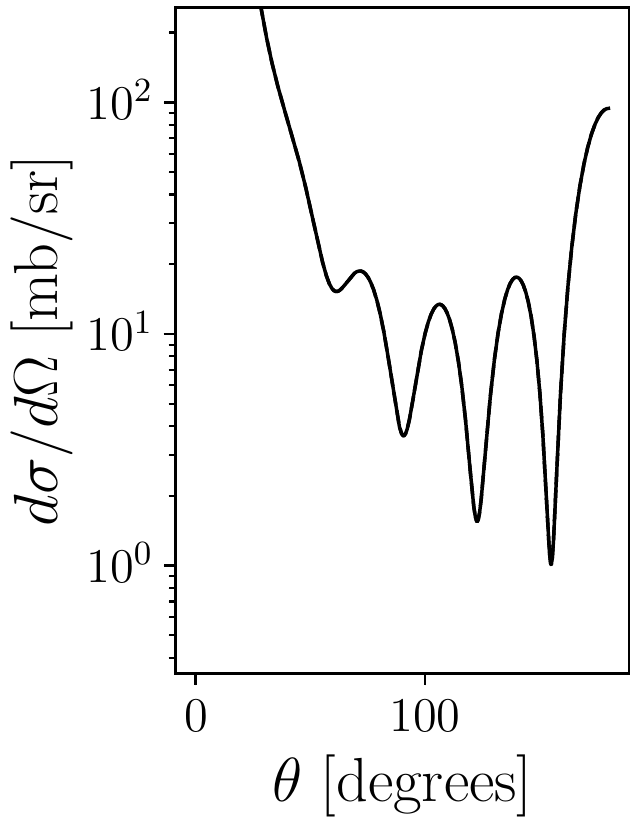}
            \caption{$E=9.5$ MeV,\\ $l=5$ angular distribution.}
           \label{fig:9.5,5AD}
        \end{figure}
        \vspace{-0.1cm}
        \begin{figure}[H]
        \hspace{-0.35cm}
            \hspace{-1cm}
            \centering  \includegraphics[width=4cm]{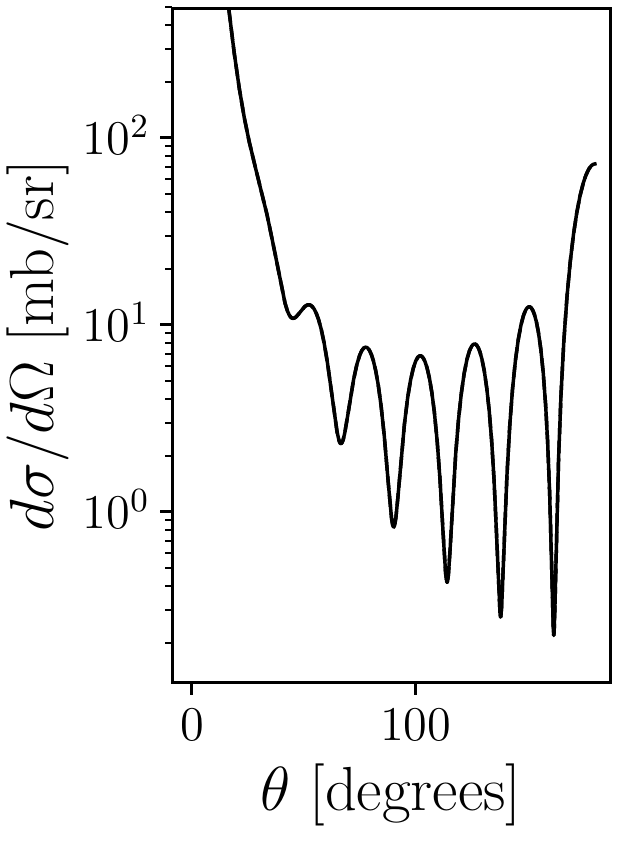}
            \caption{$E=20$ MeV,\\
            $l=7$ angular distribution.} \label{fig:20,7AD}
        \end{figure}
    \end{minipage}
\end{figure}

\begin{figure}[h!]
    \centering
    \begin{minipage}{0.35\textwidth}
        \centering
        \includegraphics[width=0.9\textwidth]{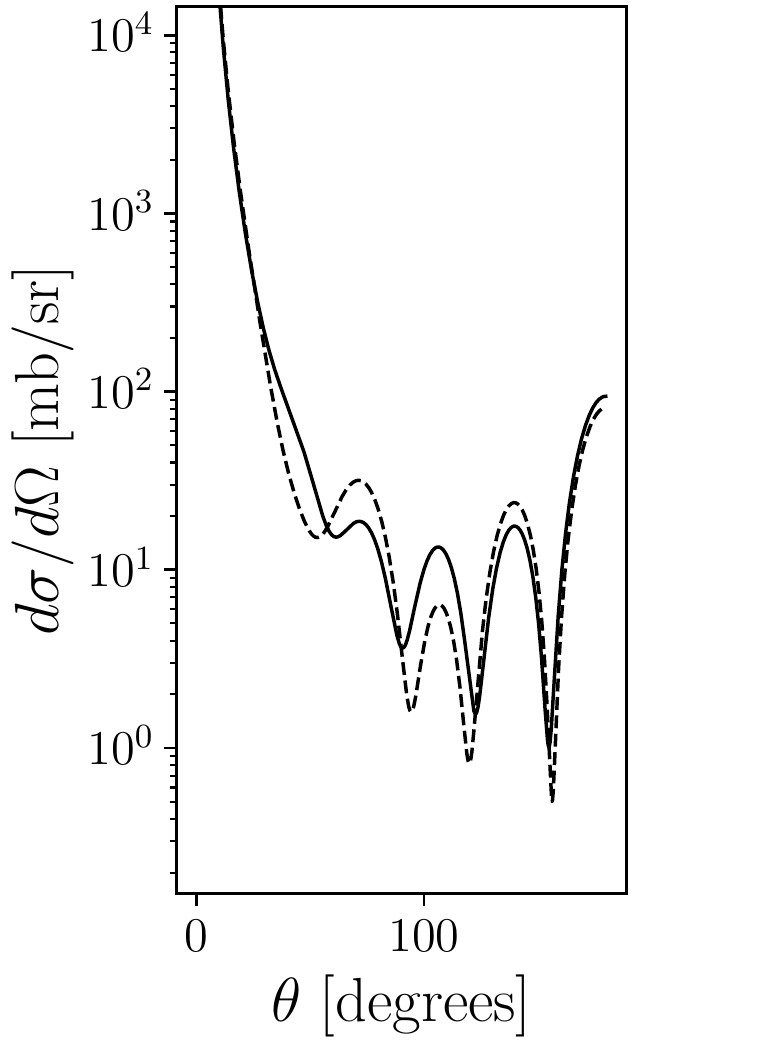}
        \caption{$E=9.5$ MeV, $l=5$ angular distribution with half-imaginary and half-real components of $N_r$.}
        \label{fig:9.5imag_real}
    \end{minipage}%
    \hspace{1cm}
    \begin{minipage}{0.35\textwidth}
        \vspace{2.6mm}
        \centering
        \includegraphics[width=0.9\textwidth]{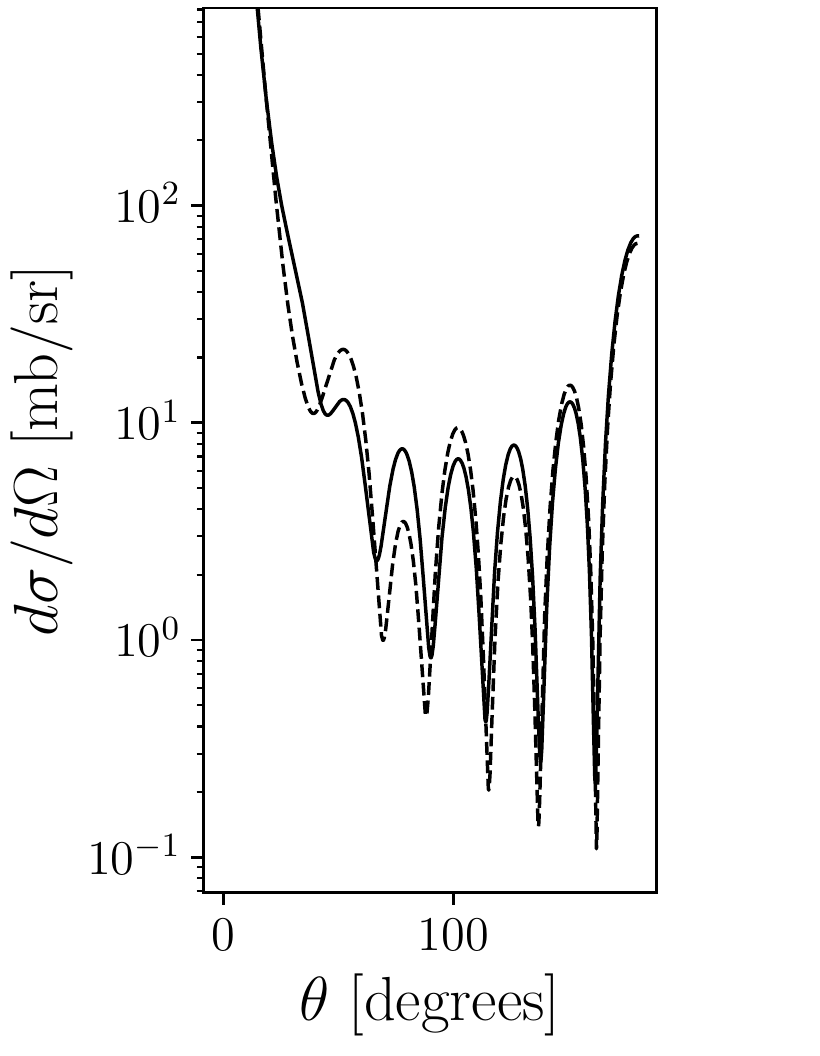}
        \vspace{0.3mm}
        \caption{$E=20$ MeV, $l=7$ angular distribution with half-imaginary and half-real components of $N_r$.}
        \label{fig:20imag_real}
    \end{minipage}
\end{figure}

\begin{figure}[h!]
    \centering
    \begin{minipage}{0.35\textwidth}
        \centering
        \includegraphics[width=0.9\textwidth]{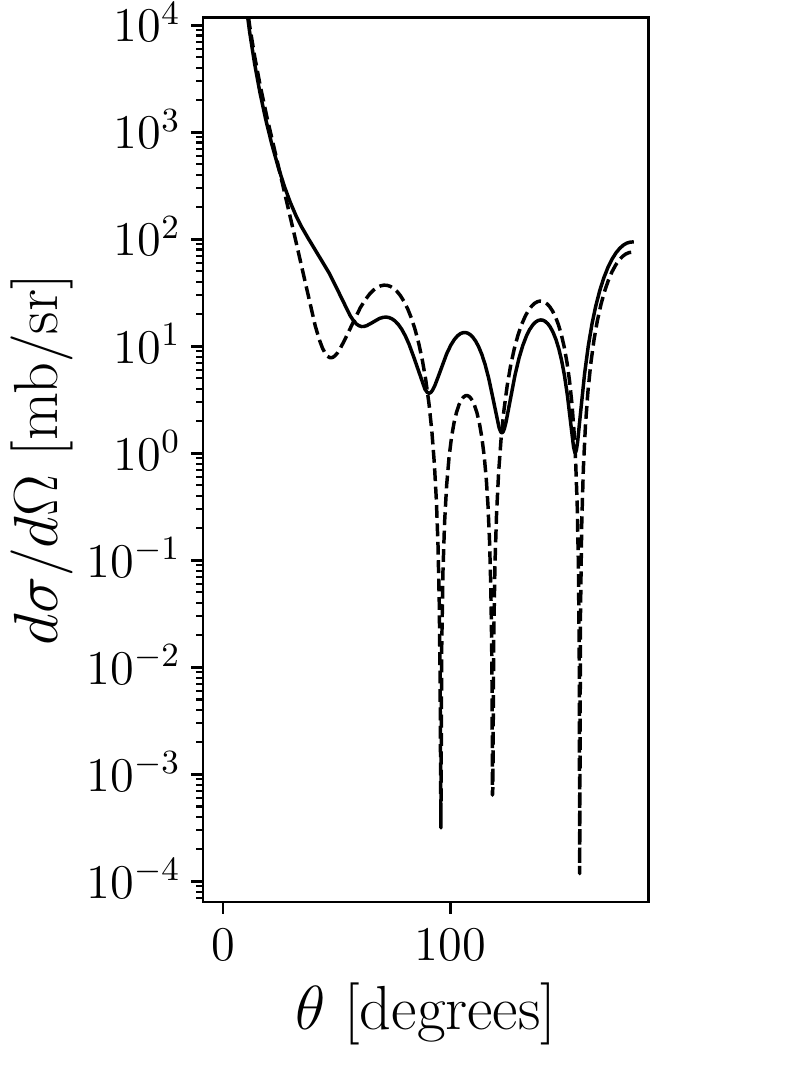}
        \caption{$E=9.5$ MeV, $l=5$ angular distribution with completely real $N_r$.}
        \label{fig:9.5real}
    \end{minipage}%
    \hspace{1cm}
    \begin{minipage}{0.35\textwidth}
        \vspace{1.2mm}
        \centering
        \includegraphics[width=0.9\textwidth]{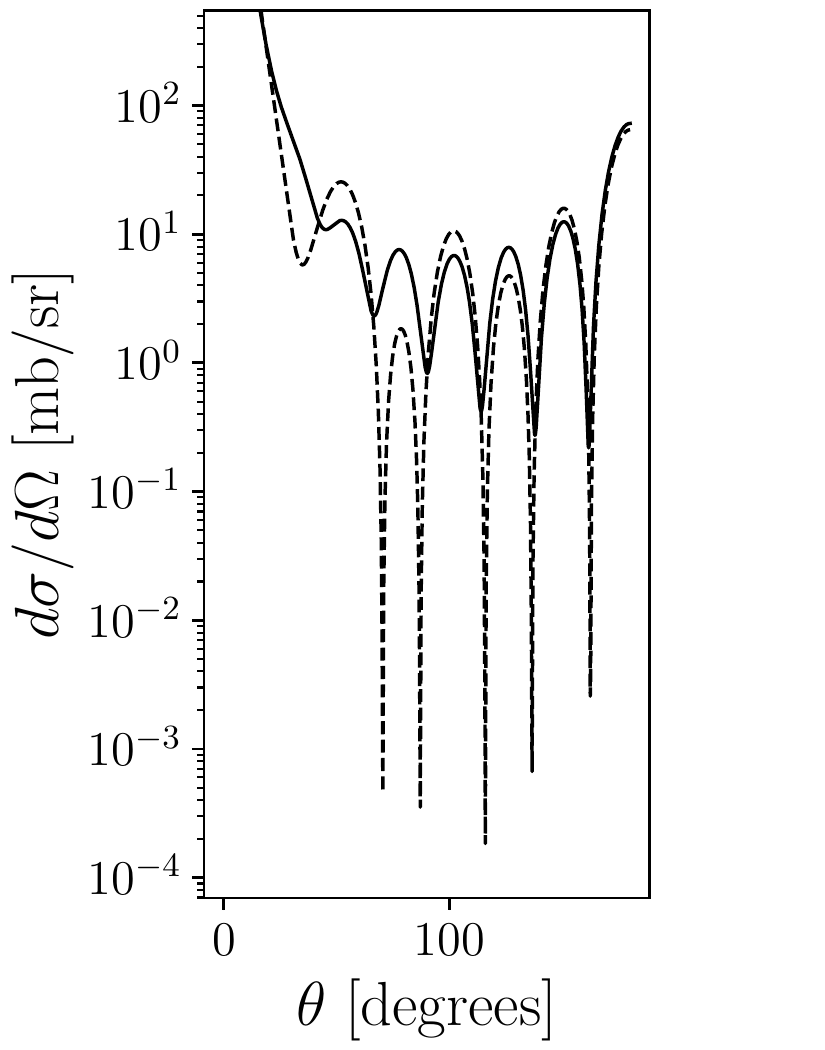}
        \vspace{1mm}
        \caption{$E=20$ MeV, $l=7$ angular distribution with completely real $N_r$.}
        \label{fig:20real}
    \end{minipage}
\end{figure}

Figures \ref{fig:9.5,5AD} and \ref{fig:20,7AD} show order-of-magnitude values in comparison to the Taylor data. Considering different real and imaginary values of the normalization constant (\ref{eqn:norm}), the distribution amplitudes take a variable range (Figures \ref{fig:9.5imag_real}, \ref{fig:20imag_real}, \ref{fig:9.5real}, and \ref{fig:20real}). In each case, the solid line indicates the classical condition, whereby there is no interference. This happens to be the case that $N_r$ is completely imaginary, since the complex conjugate of an imaginary term becomes negative, cancelling out the the cross terms:
\begin{align}
\label{eqn:angular_total2}
    \frac{d\sigma}{d\Omega}=|f_C|^2 + |f_N|^2 = \left|\frac{Z_{\text{Be}}\,Z_{\text{He}}\, \alpha \, \hslash c}{4E \sin^2{\left(\sfrac{\theta}{2}\right)}}\right|^2 + \left|N_{r} \, Y_{l}^0(\theta) \right|^2.
\end{align}
The mixed values of real and imaginary components ($\lambda = \frac{1}{2} \omega$) give a more reasonable range of amplitudes than do the completely real values. For more accurate results, the favorable magnitude of the imaginary component seems to suggest errors arising from absent phase change. The optical model used by Taylor uses the Woods-Saxon potential (\ref{eqn:AW}) for the imaginary part of the potential and the \pmb{$I \!\cdot \!l$} spin-coupling term (\ref{eqn:OM}) for back-angle corrections \cite{Aygun}.

\begin{figure}[h!]
    \vspace{-1mm}
    \centering
    \begin{minipage}{0.45\textwidth}
        \hspace*{-2mm}
        \centering
        \includegraphics[width=0.95\textwidth]{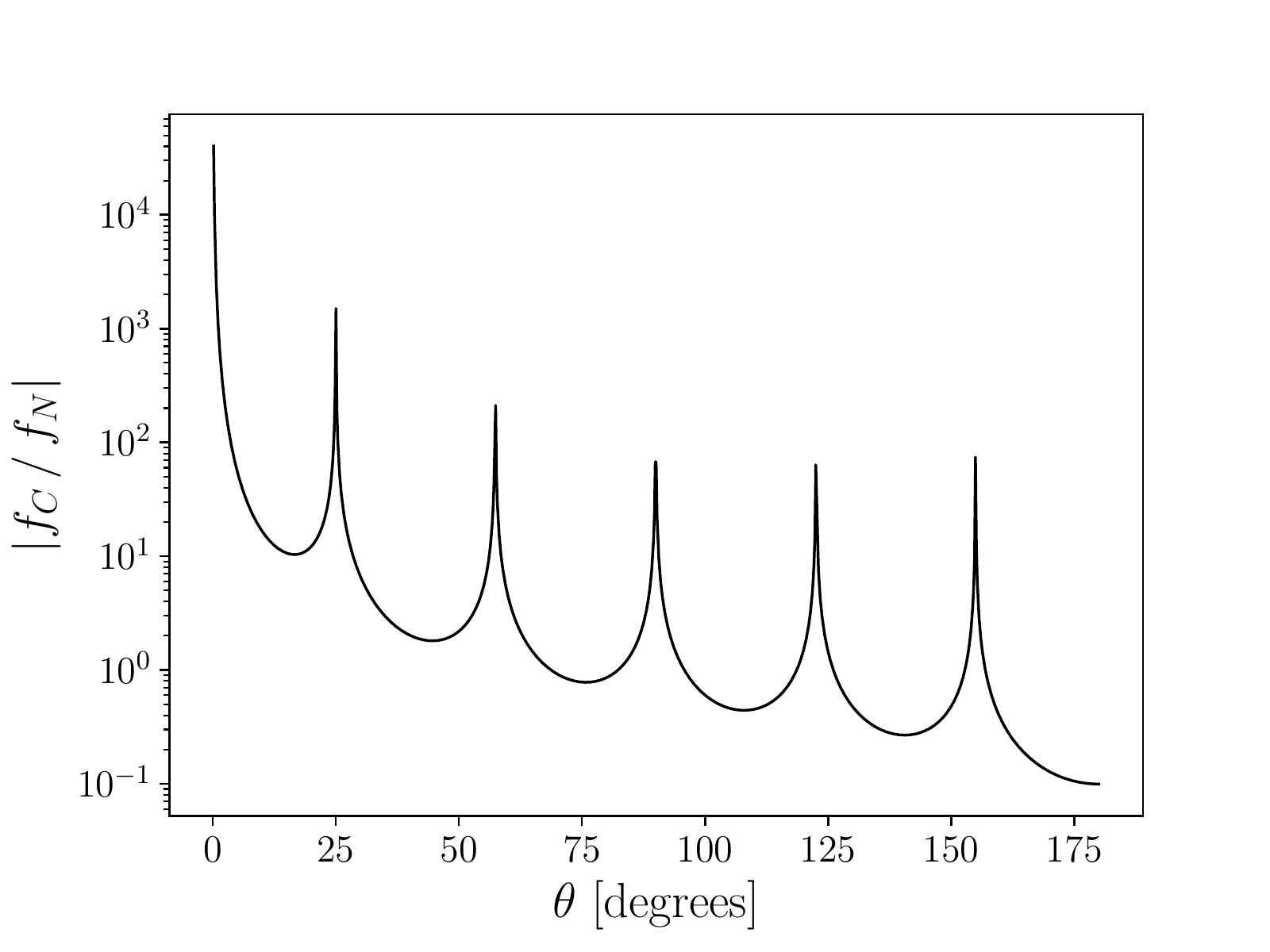}
        \caption{$E=9.5$ MeV, $l=5$ ratio of Coulomb to Nuclear processes as a function of scattering angle.}
       \label{fig:9.5ratio}
    \end{minipage}%
    \hspace{1cm}
    \begin{minipage}{0.45\textwidth}
        \hspace*{-1mm}
        \centering
        \includegraphics[width=0.95\textwidth]{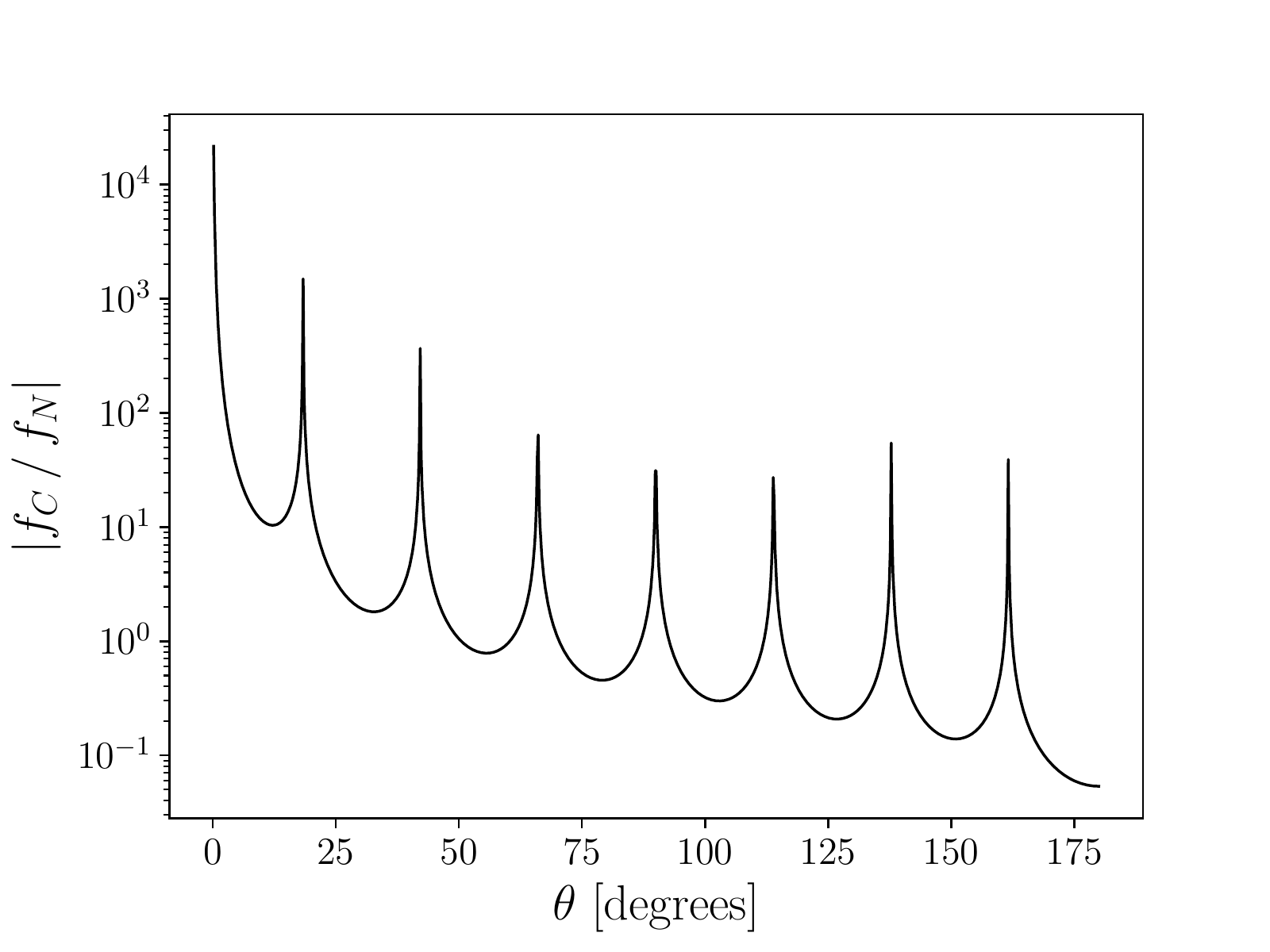}
        \caption{$E=20$ MeV, $l=7$ ratio of Coulomb to Nuclear processes as a function of scattering angle.}
        \label{fig:20ratio}
    \end{minipage}
\end{figure}

The strong influence of the proximity formalism used, shows reasonable back-angle peaks for higher scattering angles. To observe the dominance of the angular momentum component at larger angles, the ratio of the Coulomb and nuclear contributions $|f_C/f_N|$ is plotted in Figures \ref{fig:9.5ratio} and \ref{fig:20ratio}.

For both energies, and without much influence by scaling the normalization constant, the strength of the nuclear component dominates the Coulomb force beyond 160 degrees. This is once more indicative of strong angular momenta processes happening in the exit channels.

\section{Conclusions}

The primary reason for using the proximity potential for a light nuclei-nuclei interaction was to better understand the sharp angular distribution rise for back-angle scattering. The optical model is lacking in this aspect, since it is not specifically designed for direct (surface) reactions, hence requiring a more sophisticated spin-coupling term. The proximity potential here excels at showing physical insight without complicated scaling.

Inaccuracy in the proximity method of calculating the angular distribution stems largely from the liquid drop model. The real nucleus is far more complex than the assumptions of the drop, and is particularly inaccurate for light nuclei with small binding energies. Indeed, there is a Fermi superfluid inside the shell that does not rotate with the outer edges, effectively lowering the moment of inertia that is classically provided by the liquid drop model. The shell correction of Myers and Swiatecki (\ref{eqn:M}) is a strong improvement from the original model but still incomplete, despite successful predictions for spherical nuclei in their ground state \cite{Ludwig}. 

Another limitation of the proximity potential application to light nuclei arises from the surface diffuseness parameter $b$, whereby the similarity to the Coulomb radius must not be too high. The impact parameter is more precisely defined as  
\begin{align}
\label{eqn:S2}
    b'=r_p + r_t +b,
\end{align}
which gives new cross section dependence for a rotating liquid drop \cite{Swi2}. In real experiment, the impact parameter takes many different values and the nuclei scatter inelastically \cite{Krane}.

The methods used in this paper would be further improved by calculating the full Hamiltonian and considering phase change, as well as the imaginary component in $|f_C|$, which would give two more orders of freedom. Rigorous treatment of phase would require using the distorted-wave Born approximation for incident and outgoing waves. This remains an area of technical strength for the optical model. The nuclear potential is important beyond the angular distribution, since the potential energy constitutes half of the system Hamiltonian function \cite{Swi2}. To develop a more complete theory, and further understand suface phenomena, the shell structure influence on the kinetic energy might be considered. The development of a semi-empirical Hamiltonian would simplify the surface-energy coefficient in the proximity potential \cite{77} and potentially yield further insight into the $^9$Be\,+$^4$He reaction and other light nuclei scattering events.

\printbibliography

\end{document}